\documentclass[epj]{svjour}
\usepackage{graphics}
\begin{document}
\title{Nucleon sigma term and quark condensate in nuclear matter}
\author{K.~Tsushima\inst{1}, K.~Saito\inst{2}, A.~W.~Thomas\inst{3} 
\and A.~Valcarce\inst{1}
}                     
%
%
\institute{Universidad de Salamanca, E-37008, Salamanca, Spain  
\and Tokyo University of Science, Noda 278-8510, Japan 
\and Jefferson Lab, 12000 Jefferson Avenue, Newport News, VA 23606, U.S.A.}
\date{Received: date / Revised version: date}
%
\abstract{
We study the bound nucleon sigma term and the quark condensate 
in nuclear matter.
In the quark-meson coupling (QMC) model the nuclear
correction to the sigma term is small and negative, i.e., it  
decelerates the decrease of the quark condensate
in nuclear matter. However, the quark condensate in nuclear matter
is controlled primarily by the scalar-isoscalar $\sigma$ field.
Compared to the leading term, it moderates 
the decrease more than that of the nuclear sigma term alone 
at densities around and larger than the normal 
nuclear matter density.
} 
\maketitle
\section{Introduction}
\label{intro}
One of the most challenging questions in nuclear physics is
whether or not the properties of nucleon and hadrons
are modified in nuclei~\cite{BR,PPNP}.
A number of pieces of evidence, such as the nuclear EMC effect~\cite{EMC},
the quenching~\cite{gaspace,KTRiska,LugA} (enhancing~\cite{gatime})
of the space (time) component of the axial coupling constant in nuclear
$\beta$ decays, the missing strength of the response functions
and the suppression of the Coulomb sum rule~\cite{Coulomb},
have stimulated investigations of this question.
In particular, to focus on the change of the internal structure of
the nucleon (hadrons) is very appealing in the light of an approach based 
on QCD.

In addition to the recent experimental evidence for the
vector meson mass shifts in nuclei~\cite{Trnka,KEK},
some important hints concerning the change of the internal structure
of a bound nucleon has been reported
in measurements of the electromagnetic form factors of a proton
in $^4$He~\cite{JLab}. The analyses suggest a reduction of the bound
proton's electric to magnetic form factor ratio, where most of
the conventional, sophisticated approaches employing the free proton
form factors~\cite{Kelly} fail to explain the observed
effect. (See Ref.~\cite{Schiavilla} for an exploration of a  
conventional mechanism which might explain at least some of the effect.
We note, however, that for the moment that approach is hampered by a lack of 
experimental data to pin down the input parameters.)
Indeed, excellent agreement with the data is achieved when the
small correction associated with the change of the internal structure 
of the bound proton (which had been
predicted~\cite{EMff} long before the experiments) is included.

Up to now, some models
such as the quark-meson coupling (QMC) model~\cite{qmc,QMCII,GT} and
Nambu-Jona-Lasinio model~\cite{Bentz}, have opened possibilities
to understand the change of the nucleon internal structure in nuclei
based on quark degrees of freedom.
However, it is important to explore other ways to study the change in the
internal structure of the bound nucleon through various
nuclear phenomena.

In this note, we study the bound nucleon sigma term,
which has a direct connection with the QCD Hamiltonian via
Feynman-Hellmann theorem~\cite{Feynman}.
First, we study the density dependence of the nuclear correction
to the sigma term in the QMC model~\cite{qmc,QMCII}.
Next, we study the influence of the nuclear correction that
arises from the sigma term on the quark condensate.

\section{Bound nucleon sigma term}
\label{sec:1}
For the ground state of uniform, nucleon density $\rho_B$
with A nucleons and volume $V$ ($\rho_B=A/V$),
the nuclear sigma term $\sigma_A$
and the bound nucleon sigma term $\sigma_N^*$
may be defined by~\cite{Jaffe,Delome,Gammal},
\begin{eqnarray}
\sigma_A &\equiv& A\sigma_N^*(\rho_B),
\nonumber\\
&=&\frac{1}{3}\sum_{a=1}^{3}
[<A(\rho_B)|[Q_5^a,[Q_5^a,H_{QCD}]]|A(\rho_B)>
\nonumber\\
& &\hspace*{15ex}-<0|[Q_5^a,[Q_5^a,H_{QCD}]]|0>],
\label{eqAsigma}\\
&=& V\rho_B 2m_q[<A(\rho_B)|\bar{q}q|A(\rho_B)>-<0|\bar{q}q|0>],
\label{eqbsigma1}
\end{eqnarray}
where $Q_5^a$ is the weak axial charge with isospin $a$, and $H_{QCD}$
the QCD Hamiltonian with the current quark mass,
$m_q\equiv\frac{1}{2}(m_u+m_d)$, and
$\bar{q}q\equiv\frac{1}{2}(\bar{u}u+\bar{d}d)$.
Applying the Feynman-Hellmann theorem~\cite{Feynman},
we get:
\begin{eqnarray}
A\sigma^*_N(\rho_B)
&=& m_q\frac{\partial}{\partial m_q}
[<A(\rho_B)|H_{QCD}|A(\rho_B)> 
\nonumber\\
& &\hspace*{20ex} - <0|H_{QCD}|0>],
\label{eqbsigma2}\\
&=& V\rho_B m_q \frac{\partial}{\partial m_q}
\left[m_N(m_q) + \varepsilon_B(m_q,\rho_B)\right],
\label{eqbsigma3}\\
&\equiv& A[\sigma_{N free} + \delta\sigma_N^*(\rho_B)],
\label{eqbsigma4}
\end{eqnarray}
where $\sigma_{N free}\equiv m_q \frac{\partial m_N}{\partial m_q}$ is
the free nucleon sigma term (empirical value $\simeq 45$
MeV~\cite{Gasser}),
$\delta\sigma_N^*(\rho_B)\equiv m_q
\frac{\partial\varepsilon_B(m_q,\rho_B)}{\partial m_q}$
the nuclear correction to the sigma term,
and $\varepsilon_B(m_q,\rho_B)$ the
binding energy per nucleon.
At present, $\varepsilon_B(m_q,\rho_B)$ which contains the contributions
from the nucleon kinetic energy and N-N interaction etc.,
can only be calculated in a model dependent way.
Note that our focus concerns the dependence on the
current quark mass $m_q$~\cite{Feynman},
which has a direct connection with the QCD Hamiltonian through
Eqs.~(\ref{eqbsigma1})-(\ref{eqbsigma4}).
It may be contrasted with the calculation of the quark condensate
in nuclear matter~\cite{Cohen,Li,Brockmann,QMCIIqcon},
which will also be studied later.
For later convenience we recall the saturation conditions
for nuclear matter:
\begin{eqnarray}
\left.
\frac{\partial\varepsilon_B(m_q,\rho_B)}
{\partial\rho_B}\right|_{\rho_B=\rho_0}&=&0,
\label{eqsaturation1}\\
\varepsilon_B(m_q,\rho_B=\rho_0)&=&-15.7 {\rm\quad MeV}.
\label{eqsaturation2}
\end{eqnarray}

Now we study the density dependence of the 
nuclear correction to the nucleon sigma term
$\delta\sigma_N^*(\rho_B)$ in Eq.~(\ref{eqbsigma4}).
Generally, models for nuclear matter based
on quark degrees of freedom contain coupling
constants, $g_j(m_q) (j=1,2,...)$, which depend on the current
quark mass $m_q$. These coupling constants are determined so
as to satisfy the saturation conditions
Eqs.~(\ref{eqsaturation1}) and~(\ref{eqsaturation2})  
for a chosen value of $m_q$.
This also fixes the function $\varepsilon(m_q,\rho_B)$
as a function of $m_q$. Of course, since in nature we have only one 
set of quark masses, which lead to nuclear saturation at the correct place, 
it is not appropriate to include the variation of $g_j$ with $m_q$ in 
calculating the nuclear sigma term.

In the following we use different versions of the QMC model
(QMC-I~\cite{qmc}, and QMC-II~\cite{QMCII} with parameter set B),
and study the density dependence of the nuclear
correction $\delta\sigma_N^*(\rho_B)$.
We use the standard parameters, $m_q=5$ MeV, and the free nucleon bag radius
$R_N=0.8$ fm (all the corresponding parameters, values for the
coupling constants  can be found in Refs.~\cite{PPNP,qmc,QMCII}).
Thus, we suppose that $m_q=5$ MeV is the value in the QCD Hamiltonian.

\begin{figure}
\rotatebox{-90}{
\resizebox{0.35\textwidth}{!}{\includegraphics{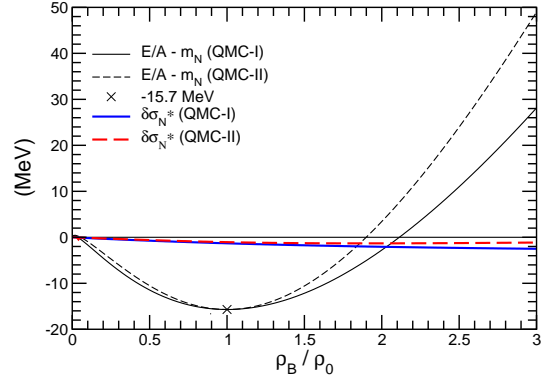}}
}
\caption{Nuclear correction to the bound nucleon
sigma term ($\delta\sigma_N^*$),
and the binding energy per nucleon
($\varepsilon_B=E/A-m_N$), where $\rho_0=0.15$ fm$^{-3}$.}
\label{figbsigma}       
\end{figure}

First, we show in Fig.~\ref{figbsigma} the nuclear correction to the
sigma term, $\delta\sigma^*_N(\rho_B)$,
together with the binding energy per nucleon, $\varepsilon_B=E/A-m_N$
(saturation curve) calculated in QMC-I and QMC-II.
%
The negative contribution of the nuclear correction to the sigma term
agrees with the result~\cite{Gammal} based on the 2-body N-N
interaction using the Skyrme model, although
the correction is small in the present case.
In Refs.~\cite{Li,Brockmann}, an effective
density dependent sigma term was introduced. Although they
estimated the $m_q$  dependence of the meson masses appearing
in the N-N interaction, in order to calculate the in-medium quark condensate,
the calculation was not based on QCD but instead was rather qualitative.
We do not attempt to make such an estimate for
the variation of the masses of the meson fields.

\section{Quark condensate in nuclear matter}
\label{sec:2}
Next, we discuss the quark condensate in nuclear matter.
The quark condensate and the nuclear sigma term are related through
Eqs.~(\ref{eqbsigma2})~-~(\ref{eqbsigma4}). Using
the Gell-Mann-Oakes-Renner (GOR) relation,
the nuclear matter to vacuum quark condensate ratio is given by:
\begin{eqnarray}
\frac{Q^*}{Q_0}
&\equiv& \frac{<\rho_B|\bar{q}q|\rho_B>}{<0|\bar{q}q|0>}
\simeq 1 -
\frac{\sigma_{N free}+\delta\sigma_N^*(\rho_B)}{m_\pi^2 f_\pi^2}\rho_B.
\label{eqqcon}
\end{eqnarray}
Without $\delta\sigma_N^*(\rho_B)$ above,
it is the usual leading term result.

\begin{figure}
\rotatebox{-90}{
\resizebox{0.35\textwidth}{!}{\includegraphics{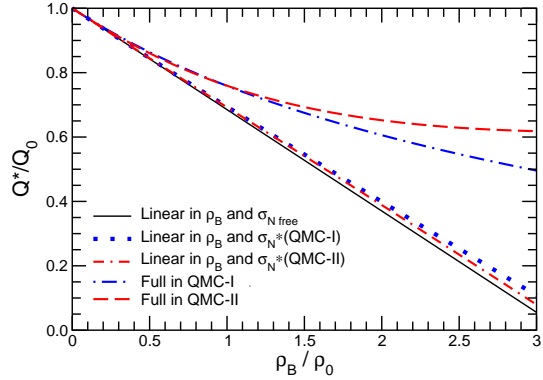}}
}
\caption{The nuclear matter to vacuum quark condensate ratio.
All results are obtained with $\sigma_{N free}=45$ MeV~\cite{Gasser}.}
\label{figqcon}       
\end{figure}

We show in Fig.~\ref{figqcon} the nuclear matter to vacuum quark
condensate ratio calculated via Eq.~(\ref{eqqcon}).
%
The solid line is the leading term with the free nucleon
sigma term, $\sigma_{N free}=45$ MeV~\cite{Gasser}, while the dotted and
dash-dash-dotted lines are those calculated with the bound nucleon
sigma term $\sigma_N^*$ in QMC-I and QMC-II.
As expected from the results for the bound nucleon sigma term,
the influence of $\delta\sigma_N^*$ to the
ratio is very small. (The full results will be discussed below.)

In the QMC model, nucleons interact through the
self-consistent exchanges of the scalar-isoscalar ($\sigma$),
vector-isoscalar ($\omega$), and vector-isovector ($\rho$) meson
mean fields, which couple directly to the quarks inside the nucleons.
In principle, although the masses of the meson fields in QMC will 
also depend on the current quark mass $m_q$, this is not a feature
of the model. Thus, in the QMC model the other nuclear correction
to the quark condensate, aside from the bound nucleon sigma term,
is expected to arise from the scalar-isoscalar $\sigma$ field.
Indeed, the full expression for the
nuclear matter to vacuum quark condensate ratio
in QMC-I is given by~\cite{QMCIIqcon},
\begin{eqnarray}
\frac{Q^*}{Q_0}
= 1 - \frac{\sigma_{N free}+\delta\sigma_N^*(\rho_B)}{m_\pi^2 f_\pi^2}
\left(\frac{m_\sigma}{g_\sigma}\right)^2 (g_\sigma\sigma),
\label{eqqconfull}
\end{eqnarray}
where $\sigma$ is the mean value of the scalar-isoscalar
$\sigma$ field in nuclear matter.
For QMC-II, one may replace in Eq.~(\ref{eqqconfull})  
$m_\sigma\to m^*_\sigma$ which depends on density or $\sigma$
but not on $m_q$ explicitly.
This expression, which is obtained using the self-consistent equation
for the $\sigma$ field, may be a characteristic feature of the QMC model.
The quark condensate in nuclear matter is
directly connected to the scalar-isoscalar $\sigma$ field,
which plays an important role in describing the properties of
nuclear matter and nuclei. This may not be a trivial result.
Note that the differences between the present expression and that in
Ref.~\cite{QMCIIqcon} arise because the present one
does not contain the terms involving the derivative of the 
quark-meson coupling constants
with respect to $m_q$. 

Next, we show in Fig.~\ref{figqcon} the full results
of Eq.~(\ref{eqqconfull}) in QMC-I (the dash-dotted line)
and QMC-II (the dashed line).
In both versions of QMC, the scalar-isoscalar $\sigma$ field
largely moderates the decrease of the quark condensate relative
to the leading term.
In Refs.~\cite{Li,Brockmann}, although the $m_q$ dependence
of the exchanged mesons in the N-N interaction
was estimated qualitatively, they also got a similar, concave shape of the
density dependence for the ratio.

\section{Summary}
\label{summary}
To summarise, it is shown that the nuclear correction to the bound
nucleon sigma term is negative, and decelerates the decrease
of the quark condensate in nuclear matter.
However, the magnitude is small and so is its effect on the
quark condensate.
The quark condensate in nuclear matter is
predominantly controlled by the 
scalar-isoscalar $\sigma$ field in the QMC model.
It moderates appreciably the decrease of the condensate
relative to the leading term at densities
around and larger than the normal nuclear matter density.

\vspace{2ex}
\noindent
{\bf Acknowledgements}\\
K.T. would like to acknowledge the warm hospitality at Jefferson Lab.,
where part of the work was carried out. This work was supported 
in part by U.S. DOE Contract No. DE-AC05-06OR23177, 
under which Jefferson Science Associates 
operate Jefferson Lab.

%

\end{document}